\documentclass[%
reprint,
amsmath,amssymb,
aps,
]{revtex4-1}
\usepackage{graphicx}

\usepackage{siunitx}

\begin{document}

\title{Coherent Jetting behind a gate-defined Channel in Bilayer Graphene}

\author{Carolin Gold}
\email{cgold@phys.ethz.ch}
\affiliation{Solid State Physics Laboratory, ETH Zürich, CH-8093 Zürich, Switzerland}
\author{Angelika Knothe}
\affiliation{National Graphene Institute, University of Manchester, Manchester M13 9PL, United Kingdom}
\author{Annika Kurzmann}
\affiliation{Solid State Physics Laboratory, ETH Zürich, CH-8093 Zürich, Switzerland}
\author{Aitor Garcia-Ruiz}
\affiliation{National Graphene Institute, University of Manchester, Manchester M13 9PL, United Kingdom}
\author{Kenji Watanabe}
\affiliation{Research Center for Functional Materials, 
	National Institute for Materials Science, 1-1 Namiki, Tsukuba 305-0044, Japan}
\author{Takashi Taniguchi}
\affiliation{International Center for Materials Nanoarchitectonics,
	National Institute for Materials Science,  1-1 Namiki, Tsukuba 305-0044, Japan}
\author{Vladimir Fal’ko}
\affiliation{National Graphene Institute, University of Manchester, Manchester M13 9PL, United Kingdom}
\affiliation{Department of Physics, University of Manchester, Manchester M13 9PL, United Kingdom}
\affiliation{Henry Royce Institute, University of Manchester, Manchester M13 9PL, United Kingdom}

\author{Klaus Ensslin}
\author{Thomas Ihn}
\affiliation{Solid State Physics Laboratory, ETH Zürich, CH-8093 Zürich, Switzerland}

\date{\today}

\begin{abstract}
Graphene has evolved as a platform for quantum transport that can compete with the best and cleanest semiconductor systems. Recently, many interesting local properties of carrier transport in graphene  have been investigated by various scanning probe techniques. Here, we report on the observation of distinct electronic jets emanating from a narrow split-gate defined channel in bilayer graphene. We find that these jets, which are visible via their interference patterns, occur predominantly with an angle of $\SI{60}{^\circ}$ between each other. This observation is related to the specific bandstructure of bilayer graphene, in particular trigonal warping, which leads to a valley-dependent selection of momenta for low-energy conduction channels. This experimental observation of electron jetting has consequences for carrier transport in graphene in general as well as for devices relying on ballistic and valley selective transport.
\end{abstract}

\maketitle 


\section{Introduction}

The remarkable electronic properties of graphene-based van-der Waals heterostructures has led to a broad interest in both their possible applications as well as fundamentally new physics phenomena \cite{Geimrisegraphene2007,YankowitzvanWaalsheterostructures2019}. The possibility to electrically control the band-gap in an unprecedented manner \cite{OostingaGateinducedinsulatingstate2008,PereiraTunableQuantumDots2007,FalkoQuantuminformationchicken2007} makes bilayer graphene a promising platform for fundamental quantum electronic building blocks such as quantum dots and quantum point contacts \cite{OverwegElectrostaticallyInducedQuantum2018,BanszerusObservationSpinOrbitGap2020,EichSpinValleyStates2018,BanszerusSingleElectronDoubleQuantum2020}.

While many global properties of these systems are well studied and understood, the local behavior of electron flow through the device cannot be measured in conventional transport experiments. Scanning probe experiments excel at providing spatial information. Among such experiments, Scanning Gate Microscopy (SGM) is a powerful tool to study the electron flow in high mobility, two-dimensional electron gases at cryogenic temperatures. To this end, a voltage-biased metallic tip is scanned above the sample surface, thus capacitively and locally perturbing the potential landscape of the device. This technique was used to image a plethora of local phenomena \cite{BraemScanninggatemicroscopy2018,AidalaImagingmagneticfocusing2007, BhandariImagingCyclotronOrbits2016,PascherImagingConductanceInteger2014,BrunWignerKondophysics2014,BrunImagingDiracfermions2019}, its milestone-experiment being the observation of branched electron flow in the two-dimensional electron gas behind quantum point contacts in Ga(Al)As-heterostructures \cite{TopinkaImagingCoherentElectron2000,TopinkaCoherentbranchedflow2001,JuraUnexpectedfeaturesbranched2007}.

Crucial requirements for observing electron jetting are (i) a tunable electrostatically defined electronic constriction avoiding edge disorder and (ii) a clean bulk allowing for ballistic transport over micrometer distances.
Both of these were realized in bilayer graphene only recently with the experimental demonstration of high-quality quantum point contacts \cite{OverwegElectrostaticallyInducedQuantum2018,BanszerusObservationSpinOrbitGap2020} and quantum dots \cite{EichSpinValleyStates2018,BanszerusSingleElectronDoubleQuantum2020}.

We use scanning gate microscopy to explore the modification of electronic transport in bilayer graphene in close vicinity to a gate-defined constriction. The thus obtained conductance maps depict two distinct jets, which emanate from the constriction at an angle of $\SI{60}{\degree}$ with respect to each other. The angular distribution of these two jets is a result of the preferential selection of momenta due to the lowest-energy modes of the channel and the trigonally warped bulk dispersion of the gapped bilayer graphene relevant for ballistic electron motion.
This is in stark contrast to the small-angle scattering origin of electron branching observed in similar measurements on Ga(Al)As heterostructures \cite{TopinkaCoherentbranchedflow2001,JuraUnexpectedfeaturesbranched2007}.

The electron jetting along preferred crystallographic directions observed in our measurements is of general relevance for carrier transport in bilayer graphene devices. Quantum devices utilizing the valley degree of freedom will be affected by these fundamental effects  and exploiting them opens new avenues for device design.

\section{SGM on a Bilayer Graphene Channel}
\label{sec: charact_no_tip}

\begin{figure*}
	\includegraphics[scale=1]{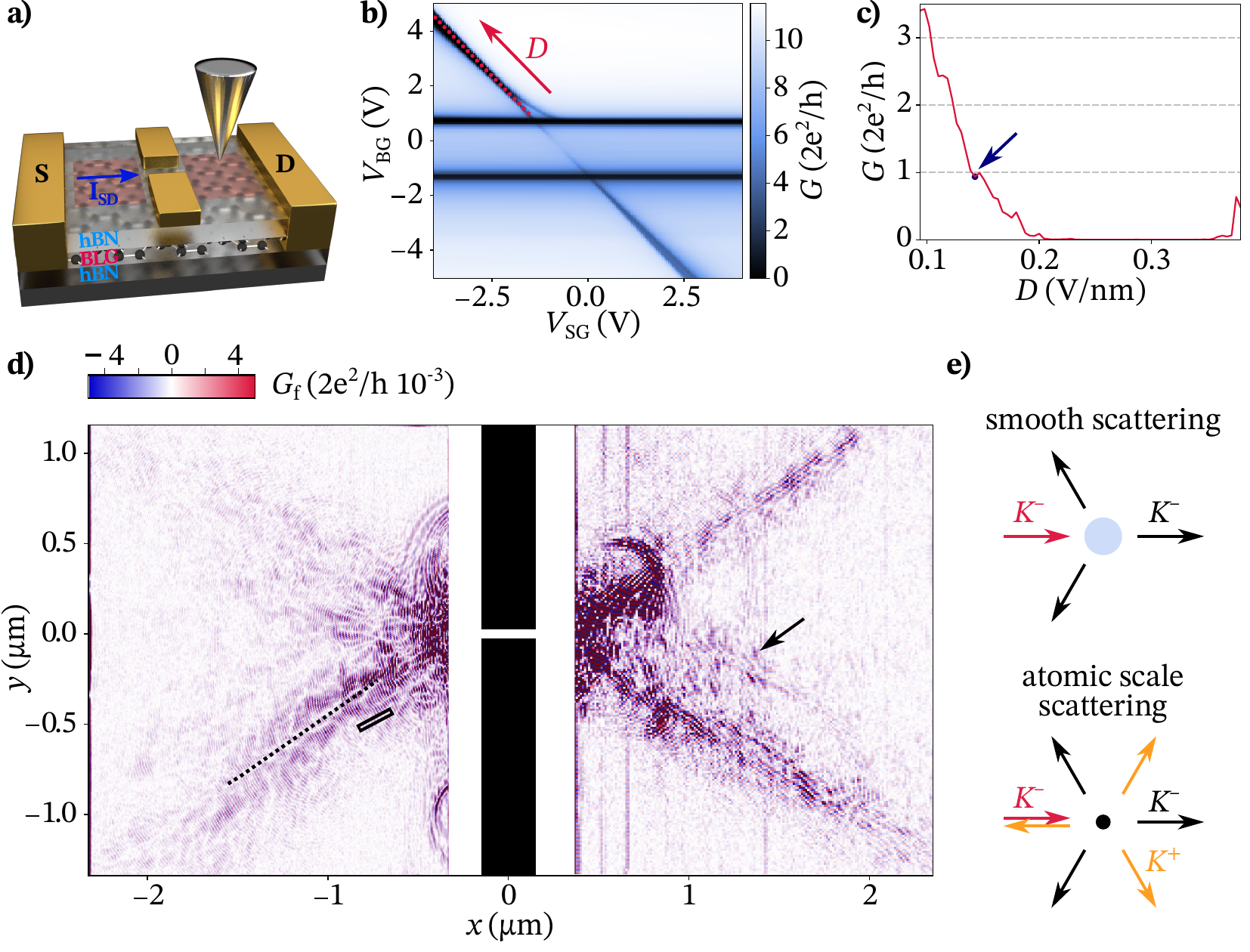}
	\caption{Observing jetting in a nano-structured bilayer graphene sample
		(a) Schematic of the measurement setup consisting of a split-gate defined channel in an encapsulated bilayer graphene (BLG) sample and a metallic atomic force microscopy tip scanned about $\SI{30}{nm}$ above the hBN surface. We measure the conductance $G(x,y)$ between source (S) and drain (D) as a function of the tip position (x,y) in the areas behind the channel (red rectangles).
		(b) Two-terminal conductance $G(V_\mathrm{SG},V_\mathrm{BG})$ as a function of split-gate ($V_\mathrm{SG}$) and back-gate ($V_\mathrm{BG}$) voltage. Charge neutrality (CNP) underneath the split-gates and hence the formation of a channel is achieved along the diagonal conductance minimum.
		(c) Conductance as a function of the displacement field $D$ along the CNP (cf red dotted line).
		(d) Filtered conductance $G_\mathrm{f}(x,y)$ on both sides of the channel (see supplementary for details on the filtering process). Clear jets with interference patterns emanate from the channel predominantly at $\SI{60}{\degree}$ with respect to each other on both sides of the channel. The black arrow marks an example of a secondary jet.
		e) Schematic of the basic scattering processes possible in the sample. Upper scheme: Valley-conserving scattering off a smooth scattering potential. Lower scheme: both valley conserving (black arrows) and valley-altering (orange arrows) scattering off the sample boundary or atomic scale defects.
	}
	\label{fig: Fig1}
\end{figure*}


\subsection{Experimental Details}
Our measurement setup consists of a sample made of a nano-structured van der Waals stack and a metallic SGM-tip scanning above it, as depicted schematically in Fig. 1a. The stack consists of a bilayer graphene (BLG) flake encapsulated between $\SI{32(40)}{nm}$ thick hexagonal Boron Nitride (hBN). Fabricated as in Ref. \cite{OverwegElectrostaticallyInducedQuantum2018}, the BLG is contacted by source and drain contacts and gated by a graphite backgate and two metallic split-gates on top of the device ($\SI{300}{nm}$ long and $\SI{50}{nm}$ apart). 
The joint action of the split-gate voltage $V_\mathrm{SG}$ applied to both split-gates and the back-gate voltage $V_\mathrm{BG}$ results in a gate-induced displacement field ${D=[C_\mathrm{BG}(V_\mathrm{BG}-V_\mathrm{BG}^0)-C_\mathrm{SG}(V_\mathrm{SG}-V_\mathrm{SG}^0)]/(2\epsilon_0)}$ in the split-gated area. Here, $C_\mathrm{SG}$ ($C_\mathrm{BG}$) are the respective split(back)-gate capacitances per unit area  \cite{ZhangDirectobservationwidely2009} and $V_\mathrm{SG}^{0}$ ($V_\mathrm{BG}^{0}$) are the offsets of the charge neutrality point in the double gated regions from zero. For appropriate gate-voltages, the resulting displacement field opens a bandgap in the double gated region while simultaneously keeping the Fermi energy in the gap. This renders the bilayer graphene insulating in the area below the split gates and confines the source-drain-current to the non-insulating channel between them (see \cite{OostingaGateinducedinsulatingstate2008} for more details).

We use SGM to study the spatial structure of electron flow in the graphene bulk behind the channel at $T=\SI{270}{mK}$. The voltage-biased, metallic scanning force microscope tip is scanned $\SI{30}{nm}$ above the surface of the top hBN. Applying a source-drain voltage $V_\mathrm{SD}=\SI{50}{\mu V}$, we measure the linear conductance $G(x,y)=I_\mathrm{SD}(x,y)/V_\mathrm{SD}$ as a function of tip position $(x,y)$ in the areas behind the channel (cf the red-marked areas in Fig. \ref{fig: Fig1}a). Due to the capacitive tip-sample coupling, the tip-voltage influences the charge density and the energy gap of the BLG locally. If the region below the tip is rendered insulating, it scatters electrons. Unless stated otherwise, all our measurements are performed at a tip-voltage $V_\mathrm{tip}=\SI{-10}{V}$ which we expect to induce a depletion disk below the tip apex (for details see supplementary S1).

\subsection{Forming a Bilayer Graphene Channel}
In order to characterize the interplay of the split-gate and the back-gate voltages, we measure the conductance $G(V_\mathrm{BG},V_\mathrm{SG})$ shown in Fig. \ref{fig: Fig1}b in the absence of the tip. We observe two conductance minima which occur at finite back-gate voltages and are independent of the split-gate voltage. These minima arise when the back-gate voltage tunes the Fermi energy in the regions outside the split-gated area into the charge neutrality point \footnote{The upper minimum most likely arises due to the charge neutrality point of the bulk regions close to the channel. The additional, second minimum occurs due to another device in parallel to the channel, which is not depicted in the schematic in Fig. \ref{fig: Fig1}a and was grounded during the measurement.}.
In addition to these minima, we observe a conductance minimum depending linearly on both the split-gate and the back-gate voltage (cf diagonal line of reduced conductance in Fig. \ref{fig: Fig1}b, partly marked by red dots). Along this minimum, the Fermi energy in the split-gated area is tuned into the bandgap opened by the gate-induced displacement field $D$. This renders the split-gated regions insulating, confining the source-drain current to the remaining conductive region between the gates. As the displacement field induced by the split- and back-gate voltages increases for larger gate-voltage differences, the bandgap itself increases along this minimum in the direction of the red arrow labeled $D$ in Fig.~\ref{fig: Fig1}b.

The electronic channel couples to the split-gates via stray-fields only, such that the electrostatic coupling of the channel to the gates is dominated by the back-gate.
The electron density in the channel thus increases with increasing displacement field. Our observation of complete pinch-off of the channel at $D \in [\SI{0.21}{V/nm},\SI{0.35}{V/nm}]$ suggests that the channel reaches its charge neutrality point at these displacement fields. Additional data discussed in previously published work reinforces this observation \cite{GoldScanningGateMicroscopy2020}.

Measuring the conductance through the channel as a function of displacement field along the red-dotted line in Fig. \ref{fig: Fig1}b yields the data depicted in Fig. \ref{fig: Fig1}c. Apart from the complete suppression of the current for the aforementioned displacement fields, Fig. \ref{fig: Fig1}c also shows an increasing conductance for decreasing displacement fields. Since the channel's charge neutrality point is within $D \in [\SI{0.21}{V/nm},\SI{0.35}{V/nm}]$, the Fermi-energy for $D<\SI{0.21}{V/nm}$ lies close to or even within the valence band within the channel. This renders the channel p-type whereas the bulk around the channel as well as the channel-ends are n-type.
All our SGM measurements are performed at a displacement field $D\approx \SI{0.14}{V/nm}$, for which the conductance is approximately $\SI{2}{\mathit{e^2/h}}$ (see the blue arrow and the blue point in Fig.~\ref{fig: Fig1}c).

\subsection{SGM behind the Channel}
\subsubsection{Jetting}
We perform SGM measurements in the regions indicated by red squares on both sides of the split-gates in Fig.~\ref{fig: Fig1}a. The thereby obtained conductance map is depicted in Fig.~\ref{fig: Fig1}d. In order to increase the visibility of the relevant patterns, we plot a filtered conductance $G_\mathrm{f}$ (see supplementary S2 for details of the filtering process).
The filtering procedure removes the background and eliminates high-frequency noise on length scales smaller than half the Fermi wavelength.
On both sides of the channel, the filtered conductance $G_\mathrm{f}$ features narrow jets with interference fringes emanating from the channel predominantly at an angle of $\SI{60}{\degree}$ with respect to each other. In proximity to the main jets, we observe secondary jets as indicated exemplarily by the black arrow in Fig.~\ref{fig: Fig1}d.
In contrast to branched-flow measurements on Ga(Al)As-heterostructures \cite{TopinkaImagingCoherentElectron2000,TopinkaCoherentbranchedflow2001}, we observe no significant reduction of the conductance along the jets modulated by interference patterns. 

\begin{figure*}
	\includegraphics[scale=1]{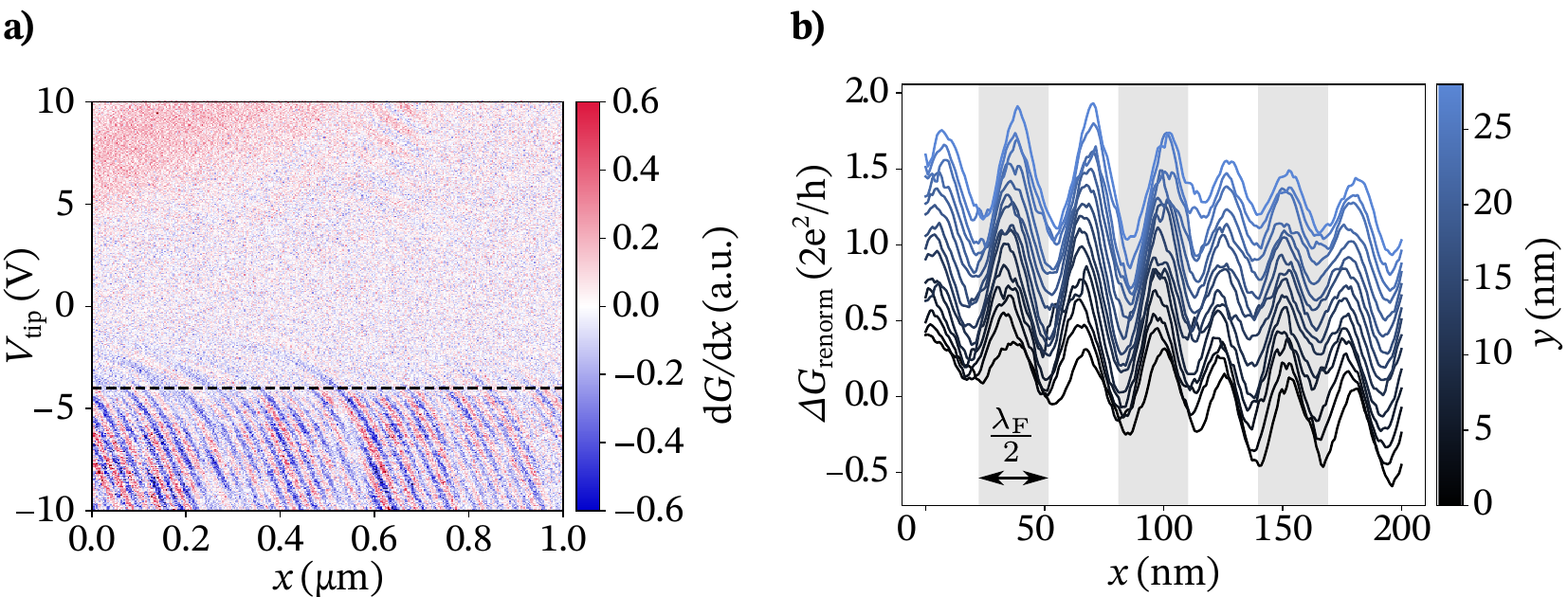}
	\caption{Interference along the jets
		(a) numerical derivative of the conductance $G(x,V_\mathrm{tip})$ along the black dotted line on the left-hand side of the channel in Fig. \ref{fig: Fig1}d. The black dashed line denotes the approximate, negative tip-voltage threshold beyond which interferences are observed.
		(b) re-normalized conductance ${\Delta G(x,y)_\mathrm{renorm}}$ (for details see supplementary) measured in the area denoted by the black rectangle on the left-hand side of the channel in Fig. \ref{fig: Fig1}d. Each line is vertically offset with respect to the previous line. A periodicity of $\lambda_\mathrm{F}/2\approx \SI{29}{nm}$ is observed (cf black arrow).
	}
	\label{fig: Fig2}
\end{figure*}

\subsubsection{Interferences: Tip-Voltage Dependence}
In order to study the influence of the tip voltage on the occurrence of interference fringes, we measure the conductance $G(x,V_\mathrm{tip})$ as a function of the tip voltage $V_\mathrm{tip}$ and tip-position $x$ along the black dashed line on the left hand side of the channel in Fig. \ref{fig: Fig1}d. The numerical derivative $dG(x,V_\mathrm{tip})/dx$ is depicted in Fig. \ref{fig: Fig2}a. We observe interference fringes for $V_\mathrm{tip}<\SI{-4}{V}$ far below the least-invasive tip voltage $V_\mathrm{tip} \approx \SI{0.25}{V}$ (for details see supplementary S3). These interference patterns almost completely disappear at $V_\mathrm{tip}\geq\SI{-4}{V}$ (cf above the dashed black line in Fig. \ref{fig: Fig2}a) and do not reappear for any more positive tip-voltages. 

The observed tip-voltage dependence suggests that the interference pattern originates from scattering of electron waves off a tip-induced depletion disk. For tip voltages more negative than the least-invasive tip voltage $V_\mathrm{tip}^{li}$, the tip reduces the charge carrier density below its apex. Once the tip voltage reaches $V_\mathrm{tip}<\SI{-4}{V}$, the Fermi-energy underneath the tip is tuned into the bandgap opened by the displacement field between back-gate and tip. This induces an insulating depletion disk below the tip. We do not expect to reach the regime in which the depletion disk fills with a p-type region for the boundary conditions of our experiment (see supplementary).

\subsubsection*{Interferences: Periodicity}
Both measurements behind the channel (cf. Fig.~\ref{fig: Fig1}d and Fig.~\ref{fig: Fig2}a) depict quasi-periodic interference fringes. In order to verify the periodicity of these patterns, we perform high-resolution scanning gate imaging in the area denoted by the black rectangle on the left hand side of the channel in Fig. \ref{fig: Fig1}d. The re-normalized difference of the conductance, $\Delta G_\mathrm{renorm}(x,y)$, measured within this area is depicted in Fig. \ref{fig: Fig2}b. Here, $x$ denotes the axis along the long side of the rectangle in Fig.~\ref{fig: Fig1}d and $y$ the axis along the short side of the rectangle. Details on the renormalization of the conductance can be found in the supplementary material (S4).
For better visibility, each line depicted in Fig.~\ref{fig: Fig2}b is slightly offset with respect to the previous line. Highlighted by the alternation of the grey/white background color, the measurement indeed reveals a periodicity of $\approx \SI{29}{nm}$. This value is in good agreement with the Fermi-wavelength estimated via a capacitive model for the van der Waals stack, for which $n_\mathrm{s}=\epsilon_0\epsilon_\mathrm{r} \left[V_\mathrm{BG}-V_\mathrm{BG}^{\mathrm{CNP}}\right]/(e d)=\SI{4.12e11}{cm^{-2}}$ and therefore $\lambda_\mathrm{F}/2\approx\SI{28}{nm}$. Here, $V_\mathrm{BG}^\mathrm{CNP}$ is the experimentally observed offset of the charge neutrality point of the bulk regions close to the channel.

\subsubsection{Magnetic Field Dependence}

\begin{figure*}
	\includegraphics[scale=1]{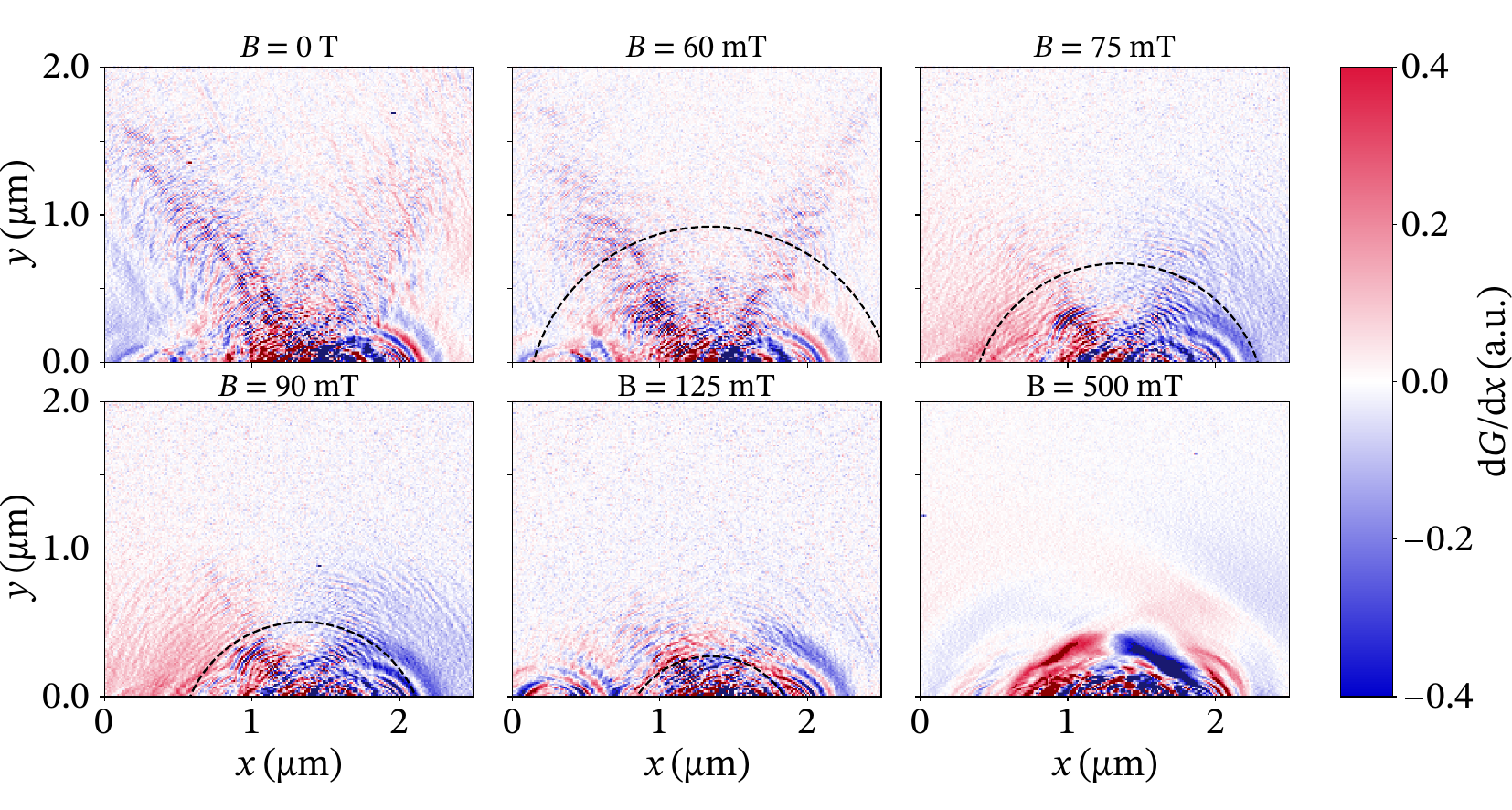}
	\caption{SGM measurements on the left-hand side of the channel in Fig. \ref{fig: Fig1}d at varying magnetic field. The respective cyclotron orbit for each magnetic field is denoted by the dashed, black half-circle centered around the channel center. For $B\neq 0$, the strongest interference patterns are observed for tip-channel distances smaller than the cyclotron orbit radius.}
	\label{fig: Fig3}
\end{figure*}

In order to shed light onto the origin and scale of the interference fringes, we measure the conductance on the left side of the channel at various small, vertically applied magnetic fields. The numerical derivative $dG(x,y)/dx$ of the measured SGM images is depicted in Fig.~\ref{fig: Fig3}. The dashed half-circles in the scans at non-zero magnetic fields indicate the classical cyclotron radius $R_\mathrm{c}=\hbar k_\mathrm{F}/eB$ around the center of the channel for the respective magnetic field.
While the interference pattern extends beyond the scan-frame for $B=\SI{0}{T}$, its radial range from the channel center is already reduced for $B=\SI{60}{mT}$. Notably, the strongest interferences at $B=\SI{60}{mT}$ are observed within the circle defined by the cyclotron radius. A further reduction of the range of the interference pattern arises systematically for $B=\SI{75}{mT}$, $\SI{90}{mT}$ and $B=\SI{125}{mT}$. For all these magnetic fields, the radial distances for which the interference fringes are observed are of the order of the cyclotron radius. At very large magnetic field (example here: $B=\SI{500}{mT}$), the cyclotron orbit is so small that no interference fringes are observed any more. Only larger scale conductance modulations observed close to the channel remain. This magnetic-field dependence of the interference patterns indicates that they are related to ballistic electron transport. 

\subsection{Understanding the 60$^{\circ}$ angle between the jets}
\begin{figure*}
	\includegraphics[scale=1.1]{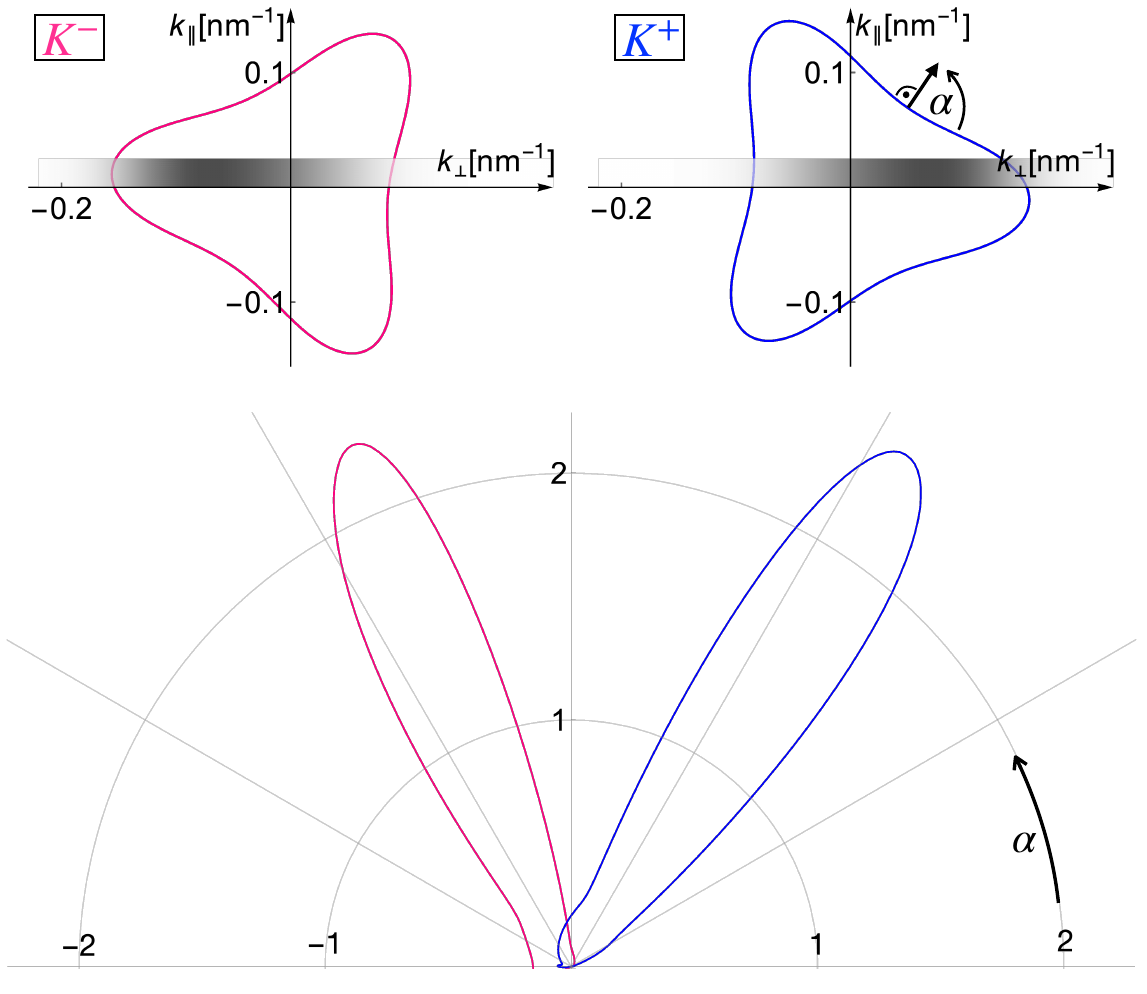}
	\caption{Fermi contours and collimated electron trajectories. Top: Fermi line for the BLG dispersion in the $K^{\pm}$ valleys in the single-gated region outside the channel for parameters $v=1.02*10^6 \text{ m/s}$, $  v_3\approx0.12 v$ and $\gamma_1=0.38\text{ eV}$. We call $k_{\parallel}$, and $k_{\perp}$, the travelling momentum along the channel and the perpendicular component, respectively. The gray bar is a density plot of the momentum distribution,  $|\Psi(k_{\perp})|^2$, for electrons in the channel, ranging from white (no occupation) to gray (substantial occupation). We compute this distribution from the lowest channel subband state, $\Psi(k_{\perp})$, in momentum space, and use it as a weight for different velocity states along the Fermi line. Bottom: combining the angular distribution of velocities obtained from the bulk dispersion in the direction of propagation with these weight factors predicts two valley-polarized jets when injecting electrons from the channel into bulk BLG for a generic orientation of the channel (here, the armchair edge of BLG is misaligned by $5^{\circ}$ with respect to the channel axis).  }
	\label{fig: Fig4}
\end{figure*}

We interpret the experimental observation of jets in scanning gate images in terms of the anisotropy of electron propagation in gapped BLG \cite{PeterfalviIntrabandelectronfocusing2012}. 
BLG's low-energy dispersion is anisotropic due to trigonal warping induced by interlayer skew hopping \cite{McCannlowenergyelectronic2007}. This effect is enhanced when one opens a gap by gating, thereby inducing an asymmetry between the layers. Any anisotropy of the dispersion entails an anisotropic angular velocity distribution of adiabatic electron trajectories. To describe this anisotropy, we use the expression  \cite{McCannlowenergyelectronic2007} 
\begin{align}
\nonumber E^{2}&= \frac{\gamma_1^2}{2}+\frac{\Delta^2}{4}+(v^2+\frac{v_3^2}{2})k^2 -\Big[\frac{(\gamma_1^2-v_3^2k^2)^2}{4}\\
&+v^2k^2[\gamma_1^2+\Delta^2+v_3^2k^2]+2\xi\gamma_1v_3v^2k^3\cos{3\varphi\Big]^{\frac{1}{2}}},
\label{eqn:BLG_disp}
\end{align}
for the electrons' low-energy dispersion in gapped BLG, obtained from the four-band tight-binding model. Here, $\xi=\pm1$ indexes the two valleys $K^{\pm}$. We use a self-consistent calculation  for the interlayer asymmetry gap, $\Delta (n_e)$, taking into account the electric susceptibility of the constituent monolayers 
and the electron density redistribution between the layers \cite{SlizovskiyDielectricsusceptibilitygraphene2019} induced by displacement field $D$ and doping (see supplementary S5.1 for details). Given  $D=\SI{143e6}{V/m}$ and the charge carrier density $n_e=\SI{4.1e11}{cm^{-2}}$ in the single-gated region outside the channel, we find $\Delta=\SI{10.3}{meV}$ and $E_\mathrm{F}= \SI{16.0}{meV}$, and the Fermi contour lines in the top panels of Fig.~\ref{fig: Fig4}. We have tested the stability of the self-consistent calculation against small variations of the hopping parameters in BLG's tight-binding description and found qualitatively similar results, namely, a bulk Fermi contour of approximately triangular shape in either valley.\\ 
For each point along the Fermi contour, the charge carrier's real space trajectory points along the direction of the normal vector as indicated in Fig.~\ref{fig: Fig4}. Approximately triangular Fermi contours like the ones shown in Fig.~\ref{fig: Fig4} hence imply significant collimation of electrons in the directions orthogonal to the dispersion's flat parts (i.e., each leg of the triangle), while only a few states are available in the directions of the triangle's corners. Consequently, any intra-valley scattering event will give rise to three scattered trajectories at mutual angles of 120$^{\circ}$, whereas direct backscattering by 180$^{\circ}$ will require a change of the valley [see schematic in  Fig.~\ref{fig: Fig1} (e)].\\ 
Similarly, we would expect six "jets" (three per valley) of electrons exiting the channel, if all states along the Fermi line were equally occupied. For isotropic injection from a point contact at an arbitrarily oriented BLG edge, three of these jets would appear in positive propagation direction (supplementary S5.1).\\ 
However, the collimation by the dispersion can be further enhanced by the electrons' momentum distribution set by the internal structure of the electron subband states in the channel from which the electrons are injected. Populating different momentum states with different weights leads to a population imbalance, selectively activating the jets. We calculate the channel spectra and states using an earlier-tested \cite{OverwegTopologicallyNontrivialValley2018, OverwegElectrostaticallyInducedQuantum2018, KnotheInfluenceminivalleysBerry2018, LaneSemimetallicfeaturesquantum2019b} numerical model for the BLG quantum wire (supplementary S5.2), where, for a channel conductance of $\sim 2e^2/h$, only the lowest channel subbands are occupied. 
Hence, we weigh the occupancy of the injected states using the Fourier transform of the lowest subband channel wave function,  $|\Psi(k_{\perp})|^2$, indicated by the grey-scale bar on the $k_{\perp}$ inset in Fig.~\ref{fig: Fig4}.

In the bottom panel of Fig.~\ref{fig: Fig4}, we show the velocity distribution for a generic orientation of the channel (e.g.,  $5^{\circ}$ misalignment between the BLG zig-zag axis and the sample edge) and including a 5$^{\circ}$ angular broadening. Two valley-polarized electronic jets emerge from the QPC at a mutual angle of $60^{\circ}$. For an arbitrary edge orientation, the jets may appear deformed, but the $60^{\circ}$ angle between them is a universal feature, prescribed by trigonal warping of a gapped BLG dispersion (see supplementary S5.3).

The anisotropic distribution and valley selectivity of the jets yields a distinct configuration of possible scattering mechanisms. The most basic of these scattering processes are depicted in Fig.~\ref{fig: Fig1}e. Scattering off a smooth scattering potential, as e.g. introduced in the graphene layer by the tip-induced electrostatic potential, results in the valley-conserving scattering depicted in the upper panel of Fig.~\ref{fig: Fig1}e. The incoming electron (red arrow), which we assume to be in the $K^-$ valley as an example, scatters off the smooth scattering potential denoted by the light-blue circle. The intra-valley scattering results in three $K^-$-valley trajectories at a mutual angle of $\SI{120}{\degree}$ (black arrows). On the contrary, scattering off the atomically rough sample boundary or an atomic scale defect, as e.g. introduced by a lattice defect, can either keep or alter the valley quantum number. As depicted in the lower panel in Fig.~\ref{fig: Fig1}e, the incoming trajectory (red arrow) in valley $K^-$ can thus be scattered into six different trajectories by an atomic scale scattering site (small black circle). While three of these trajectories (black arrows) conserve the valley quantum number, the atomic-scale scattering site can also introduce inter-valley scattering resulting in the three trajectories (orange arrows) in the other valley ($K^+$). 

As the smooth tip potential does not allow for inter-valley scattering, backscattering by $\SI{180}{\degree}$ off the tip is prohibited. This agrees with our observation that the conductance along the jets is generally not  reduced, in contrast to the measurements of branched flow behind Ga(Al)As-channels \cite{TopinkaImagingCoherentElectron2000,TopinkaCoherentbranchedflow2001,JuraUnexpectedfeaturesbranched2007} where backscattering of the tip-induced potential is the dominant mechanism. If a scattering site is positioned right within the direction of the jets emanating from the channel, scattering can give rise to secondary jets. Such secondary jets are observed in the experimental data, an example being marked by the black arrow in Fig.~\ref{fig: Fig1}d.
 
As the elastic mean free path of the charge carriers outside the channel is on the same scale as our sample size, carriers emanating from the channel will predominantly follow the directions of the jets. Scattering off atomical defects in the bulk or at the sample boundaries or the tip-induced electrostatic potential changes the direction of the carriers away from the jets. Depending on the new direction and following scattering events, the charge carriers complete different paths before reaching the drain contact or being backscattered through the channel, leading to classical conductance fluctuations. Quantum interference occurring between paths with the same valley quantum number lead to the distinct interference pattern observed along the jets in the experimental data on the scale of half the Fermi wavelength (Fig.~\ref{fig: Fig1}d, Fig.~\ref{fig: Fig2} and Fig.~\ref{fig: Fig3}). The interference stripes may have any orientation with respect to the jet direction. This is consistent with our experimental observation in graphene and in contrast with the observations in Ga(Al)As-systems \cite{TopinkaImagingCoherentElectron2000,TopinkaCoherentbranchedflow2001,JuraUnexpectedfeaturesbranched2007}.



\section{Conclusion}

We presented scanning gate microscopy measurements in close vicinity to a split-gate defined channel in high-quality bilayer graphene. The thereby obtained conductance maps depict two distinct jets emanating from the channel at an angle of $\SI{60}{\degree}$ with respect to each other. 
These jets become visible through $\lambda_\mathrm{F}/2$-periodic interference fringes which exist for tip-voltages ${V_\mathrm{tip} \in [\SI{-10}{V},\SI{-4}{V}]}$. The magnetic field dependence of these interference patterns suggests their relation to ballistic electron transport.
In contrast to similar measurements on Ga(Al)As heterostructures, the occurrence of the jets is not rooted in random electron lensing due to small-angle scattering in a correlated long-range disorder potential. Rather, the jets originate from the distinct bandstructure of bilayer graphene. The trigonal warping of the latter in conjunction with the valley-dependent selection of momenta in low-energy conducting channels results in two valley-polarized jets emerging from the channel at a mutual angle of $\SI{60}{\degree}$.

\section*{Acknowledgments}

We thank Peter Rickhaus for fruitful discussions and Peter Märki, Thomas Bähler as well as the staff of the ETH cleanroom facility FIRST for their technical support. We also acknowledge financial support by the ETH Zurich grant ETH-38 17-2 and the European Graphene Flagship. Growth of hexagonal boron nitride crystals was supported by the Elemental Strategy Initiative conducted by the MEXT, Japan, Grant Number JPMXP0112101001, JSPS KAKENHI Grant Number JP20H00354 and the CREST(JPMJCR15F3), JST. We acknowledge funding from the Core3 European Graphene Flagship Project, the European Quantum Technology Project 2D-SIPC, the ERC Synergy Grant Hetero2D, and EPSRC grants EP/S030719/1 and EP/N010345/1.

\paragraph*{Authors contributions}
C.G. and A.Ku. designed the experiment with the assistance of K.E. and T.I. C.G. performed the experiments and analyzed the data. A.Ku. fabricated the sample. K.W. and T.T. synthesized the hBN crystals. A.Kn. and A.G.-R. developed the theoretical understanding, where A.Kn. developed the model of the channel and calculated the jetting calculations, while A.G.-R. performed the self-consistent calculation of the dispersion. V.F., K.E. and T.I. supervised the project. All authors discussed the results.

\paragraph*{Data availability}
The data supporting the findings of this study will be made available via the ETH Research Collection .

\paragraph*{Competing interests}
The authors declare no competing interests. 

\bibliography{P05_Channel_Interferences_library.bib}

\begin{thebibliography}{27}%
\makeatletter
\providecommand \@ifxundefined [1]{%
 \@ifx{#1\undefined}
}%
\providecommand \@ifnum [1]{%
 \ifnum #1\expandafter \@firstoftwo
 \else \expandafter \@secondoftwo
 \fi
}%
\providecommand \@ifx [1]{%
 \ifx #1\expandafter \@firstoftwo
 \else \expandafter \@secondoftwo
 \fi
}%
\providecommand \natexlab [1]{#1}%
\providecommand \enquote  [1]{``#1''}%
\providecommand \bibnamefont  [1]{#1}%
\providecommand \bibfnamefont [1]{#1}%
\providecommand \citenamefont [1]{#1}%
\providecommand \href@noop [0]{\@secondoftwo}%
\providecommand \href [0]{\begingroup \@sanitize@url \@href}%
\providecommand \@href[1]{\@@startlink{#1}\@@href}%
\providecommand \@@href[1]{\endgroup#1\@@endlink}%
\providecommand \@sanitize@url [0]{\catcode `\\12\catcode `\$12\catcode
  `\&12\catcode `\#12\catcode `\^12\catcode `\_12\catcode `\%12\relax}%
\providecommand \@@startlink[1]{}%
\providecommand \@@endlink[0]{}%
\providecommand \url  [0]{\begingroup\@sanitize@url \@url }%
\providecommand \@url [1]{\endgroup\@href {#1}{\urlprefix }}%
\providecommand \urlprefix  [0]{URL }%
\providecommand \Eprint [0]{\href }%
\providecommand \doibase [0]{https://doi.org/}%
\providecommand \selectlanguage [0]{\@gobble}%
\providecommand \bibinfo  [0]{\@secondoftwo}%
\providecommand \bibfield  [0]{\@secondoftwo}%
\providecommand \translation [1]{[#1]}%
\providecommand \BibitemOpen [0]{}%
\providecommand \bibitemStop [0]{}%
\providecommand \bibitemNoStop [0]{.\EOS\space}%
\providecommand \EOS [0]{\spacefactor3000\relax}%
\providecommand \BibitemShut  [1]{\csname bibitem#1\endcsname}%
\let\auto@bib@innerbib\@empty
\bibitem [{\citenamefont {Geim}\ and\ \citenamefont
  {Novoselov}(2007)}]{Geimrisegraphene2007}%
  \BibitemOpen
  \bibfield  {author} {\bibinfo {author} {\bibfnamefont {A.~K.}\ \bibnamefont
  {Geim}}\ and\ \bibinfo {author} {\bibfnamefont {K.~S.}\ \bibnamefont
  {Novoselov}},\ }\bibfield  {title} {\bibinfo {title} {The rise of graphene},\
  }\href {https://doi.org/10.1038/nmat1849} {\bibfield  {journal} {\bibinfo
  {journal} {Nature Materials}\ }\textbf {\bibinfo {volume} {6}},\ \bibinfo
  {pages} {183} (\bibinfo {year} {2007})}\BibitemShut {NoStop}%
\bibitem [{\citenamefont {Yankowitz}\ \emph {et~al.}(2019)\citenamefont
  {Yankowitz}, \citenamefont {Ma}, \citenamefont {{Jarillo-Herrero}},\ and\
  \citenamefont {LeRoy}}]{YankowitzvanWaalsheterostructures2019}%
  \BibitemOpen
  \bibfield  {author} {\bibinfo {author} {\bibfnamefont {M.}~\bibnamefont
  {Yankowitz}}, \bibinfo {author} {\bibfnamefont {Q.}~\bibnamefont {Ma}},
  \bibinfo {author} {\bibfnamefont {P.}~\bibnamefont {{Jarillo-Herrero}}},\
  and\ \bibinfo {author} {\bibfnamefont {B.~J.}\ \bibnamefont {LeRoy}},\
  }\bibfield  {title} {\bibinfo {title} {Van der {{Waals}} heterostructures
  combining graphene and hexagonal boron nitride},\ }\href
  {https://doi.org/10.1038/s42254-018-0016-0} {\bibfield  {journal} {\bibinfo
  {journal} {Nature Reviews Physics}\ }\textbf {\bibinfo {volume} {1}},\
  \bibinfo {pages} {112} (\bibinfo {year} {2019})}\BibitemShut {NoStop}%
\bibitem [{\citenamefont {Oostinga}\ \emph {et~al.}(2008)\citenamefont
  {Oostinga}, \citenamefont {Heersche}, \citenamefont {Liu}, \citenamefont
  {Morpurgo},\ and\ \citenamefont
  {Vandersypen}}]{OostingaGateinducedinsulatingstate2008}%
  \BibitemOpen
  \bibfield  {author} {\bibinfo {author} {\bibfnamefont {J.~B.}\ \bibnamefont
  {Oostinga}}, \bibinfo {author} {\bibfnamefont {H.~B.}\ \bibnamefont
  {Heersche}}, \bibinfo {author} {\bibfnamefont {X.}~\bibnamefont {Liu}},
  \bibinfo {author} {\bibfnamefont {A.~F.}\ \bibnamefont {Morpurgo}},\ and\
  \bibinfo {author} {\bibfnamefont {L.~M.~K.}\ \bibnamefont {Vandersypen}},\
  }\bibfield  {title} {\bibinfo {title} {Gate-induced insulating state in
  bilayer graphene devices},\ }\href {https://doi.org/10.1038/nmat2082}
  {\bibfield  {journal} {\bibinfo  {journal} {Nature Materials}\ }\textbf
  {\bibinfo {volume} {7}},\ \bibinfo {pages} {151} (\bibinfo {year}
  {2008})}\BibitemShut {NoStop}%
\bibitem [{\citenamefont {Pereira}\ \emph {et~al.}(2007)\citenamefont
  {Pereira}, \citenamefont {Vasilopoulos},\ and\ \citenamefont
  {Peeters}}]{PereiraTunableQuantumDots2007}%
  \BibitemOpen
  \bibfield  {author} {\bibinfo {author} {\bibfnamefont {J.~M.}\ \bibnamefont
  {Pereira}}, \bibinfo {author} {\bibfnamefont {P.}~\bibnamefont
  {Vasilopoulos}},\ and\ \bibinfo {author} {\bibfnamefont {F.~M.}\ \bibnamefont
  {Peeters}},\ }\bibfield  {title} {\bibinfo {title} {Tunable {{Quantum Dots}}
  in {{Bilayer Graphene}}},\ }\href {https://doi.org/10.1021/nl062967s}
  {\bibfield  {journal} {\bibinfo  {journal} {Nano Letters}\ }\textbf {\bibinfo
  {volume} {7}},\ \bibinfo {pages} {946} (\bibinfo {year} {2007})}\BibitemShut
  {NoStop}%
\bibitem [{\citenamefont {Fal'ko}(2007)}]{FalkoQuantuminformationchicken2007}%
  \BibitemOpen
  \bibfield  {author} {\bibinfo {author} {\bibfnamefont {V.}~\bibnamefont
  {Fal'ko}},\ }\bibfield  {title} {\bibinfo {title} {Quantum information on
  chicken wire},\ }\href {https://doi.org/10.1038/nphys556} {\bibfield
  {journal} {\bibinfo  {journal} {Nature Physics}\ }\textbf {\bibinfo {volume}
  {3}},\ \bibinfo {pages} {151} (\bibinfo {year} {2007})}\BibitemShut {NoStop}%
\bibitem [{\citenamefont {Overweg}\ \emph
  {et~al.}(2018{\natexlab{a}})\citenamefont {Overweg}, \citenamefont
  {Eggimann}, \citenamefont {Chen}, \citenamefont {Slizovskiy}, \citenamefont
  {Eich}, \citenamefont {Pisoni}, \citenamefont {Lee}, \citenamefont
  {Rickhaus}, \citenamefont {Watanabe}, \citenamefont {Taniguchi},
  \citenamefont {Fal'ko}, \citenamefont {Ihn},\ and\ \citenamefont
  {Ensslin}}]{OverwegElectrostaticallyInducedQuantum2018}%
  \BibitemOpen
  \bibfield  {author} {\bibinfo {author} {\bibfnamefont {H.}~\bibnamefont
  {Overweg}}, \bibinfo {author} {\bibfnamefont {H.}~\bibnamefont {Eggimann}},
  \bibinfo {author} {\bibfnamefont {X.}~\bibnamefont {Chen}}, \bibinfo {author}
  {\bibfnamefont {S.}~\bibnamefont {Slizovskiy}}, \bibinfo {author}
  {\bibfnamefont {M.}~\bibnamefont {Eich}}, \bibinfo {author} {\bibfnamefont
  {R.}~\bibnamefont {Pisoni}}, \bibinfo {author} {\bibfnamefont
  {Y.}~\bibnamefont {Lee}}, \bibinfo {author} {\bibfnamefont {P.}~\bibnamefont
  {Rickhaus}}, \bibinfo {author} {\bibfnamefont {K.}~\bibnamefont {Watanabe}},
  \bibinfo {author} {\bibfnamefont {T.}~\bibnamefont {Taniguchi}}, \bibinfo
  {author} {\bibfnamefont {V.}~\bibnamefont {Fal'ko}}, \bibinfo {author}
  {\bibfnamefont {T.}~\bibnamefont {Ihn}},\ and\ \bibinfo {author}
  {\bibfnamefont {K.}~\bibnamefont {Ensslin}},\ }\bibfield  {title} {\bibinfo
  {title} {Electrostatically {{Induced Quantum Point Contacts}} in {{Bilayer
  Graphene}}},\ }\href {https://doi.org/10.1021/acs.nanolett.7b04666}
  {\bibfield  {journal} {\bibinfo  {journal} {Nano Letters}\ }\textbf {\bibinfo
  {volume} {18}},\ \bibinfo {pages} {553} (\bibinfo {year}
  {2018}{\natexlab{a}})}\BibitemShut {NoStop}%
\bibitem [{\citenamefont {Banszerus}\ \emph
  {et~al.}(2020{\natexlab{a}})\citenamefont {Banszerus}, \citenamefont {Frohn},
  \citenamefont {Fabian}, \citenamefont {Somanchi}, \citenamefont {Epping},
  \citenamefont {M{\"u}ller}, \citenamefont {Neumaier}, \citenamefont
  {Watanabe}, \citenamefont {Taniguchi}, \citenamefont {Libisch}, \citenamefont
  {Beschoten}, \citenamefont {Hassler},\ and\ \citenamefont
  {Stampfer}}]{BanszerusObservationSpinOrbitGap2020}%
  \BibitemOpen
  \bibfield  {author} {\bibinfo {author} {\bibfnamefont {L.}~\bibnamefont
  {Banszerus}}, \bibinfo {author} {\bibfnamefont {B.}~\bibnamefont {Frohn}},
  \bibinfo {author} {\bibfnamefont {T.}~\bibnamefont {Fabian}}, \bibinfo
  {author} {\bibfnamefont {S.}~\bibnamefont {Somanchi}}, \bibinfo {author}
  {\bibfnamefont {A.}~\bibnamefont {Epping}}, \bibinfo {author} {\bibfnamefont
  {M.}~\bibnamefont {M{\"u}ller}}, \bibinfo {author} {\bibfnamefont
  {D.}~\bibnamefont {Neumaier}}, \bibinfo {author} {\bibfnamefont
  {K.}~\bibnamefont {Watanabe}}, \bibinfo {author} {\bibfnamefont
  {T.}~\bibnamefont {Taniguchi}}, \bibinfo {author} {\bibfnamefont
  {F.}~\bibnamefont {Libisch}}, \bibinfo {author} {\bibfnamefont
  {B.}~\bibnamefont {Beschoten}}, \bibinfo {author} {\bibfnamefont
  {F.}~\bibnamefont {Hassler}},\ and\ \bibinfo {author} {\bibfnamefont
  {C.}~\bibnamefont {Stampfer}},\ }\bibfield  {title} {\bibinfo {title}
  {Observation of the {{Spin}}-{{Orbit Gap}} in {{Bilayer Graphene}} by
  {{One}}-{{Dimensional Ballistic Transport}}},\ }\href
  {https://doi.org/10.1103/PhysRevLett.124.177701} {\bibfield  {journal}
  {\bibinfo  {journal} {Physical Review Letters}\ }\textbf {\bibinfo {volume}
  {124}},\ \bibinfo {pages} {177701} (\bibinfo {year}
  {2020}{\natexlab{a}})}\BibitemShut {NoStop}%
\bibitem [{\citenamefont {Eich}\ \emph {et~al.}(2018)\citenamefont {Eich},
  \citenamefont {Pisoni}, \citenamefont {Overweg}, \citenamefont {Kurzmann},
  \citenamefont {Lee}, \citenamefont {Rickhaus}, \citenamefont {Ihn},
  \citenamefont {Ensslin}, \citenamefont {Herman}, \citenamefont {Sigrist},
  \citenamefont {Watanabe},\ and\ \citenamefont
  {Taniguchi}}]{EichSpinValleyStates2018}%
  \BibitemOpen
  \bibfield  {author} {\bibinfo {author} {\bibfnamefont {M.}~\bibnamefont
  {Eich}}, \bibinfo {author} {\bibfnamefont {R.}~\bibnamefont {Pisoni}},
  \bibinfo {author} {\bibfnamefont {H.}~\bibnamefont {Overweg}}, \bibinfo
  {author} {\bibfnamefont {A.}~\bibnamefont {Kurzmann}}, \bibinfo {author}
  {\bibfnamefont {Y.}~\bibnamefont {Lee}}, \bibinfo {author} {\bibfnamefont
  {P.}~\bibnamefont {Rickhaus}}, \bibinfo {author} {\bibfnamefont
  {T.}~\bibnamefont {Ihn}}, \bibinfo {author} {\bibfnamefont {K.}~\bibnamefont
  {Ensslin}}, \bibinfo {author} {\bibfnamefont {F.}~\bibnamefont {Herman}},
  \bibinfo {author} {\bibfnamefont {M.}~\bibnamefont {Sigrist}}, \bibinfo
  {author} {\bibfnamefont {K.}~\bibnamefont {Watanabe}},\ and\ \bibinfo
  {author} {\bibfnamefont {T.}~\bibnamefont {Taniguchi}},\ }\bibfield  {title}
  {\bibinfo {title} {Spin and {{Valley States}} in {{Gate}}-{{Defined Bilayer
  Graphene Quantum Dots}}},\ }\href {https://doi.org/10.1103/PhysRevX.8.031023}
  {\bibfield  {journal} {\bibinfo  {journal} {Physical Review X}\ }\textbf
  {\bibinfo {volume} {8}},\ \bibinfo {pages} {031023} (\bibinfo {year}
  {2018})}\BibitemShut {NoStop}%
\bibitem [{\citenamefont {Banszerus}\ \emph
  {et~al.}(2020{\natexlab{b}})\citenamefont {Banszerus}, \citenamefont
  {M{\"o}ller}, \citenamefont {Icking}, \citenamefont {Watanabe}, \citenamefont
  {Taniguchi}, \citenamefont {Volk},\ and\ \citenamefont
  {Stampfer}}]{BanszerusSingleElectronDoubleQuantum2020}%
  \BibitemOpen
  \bibfield  {author} {\bibinfo {author} {\bibfnamefont {L.}~\bibnamefont
  {Banszerus}}, \bibinfo {author} {\bibfnamefont {S.}~\bibnamefont
  {M{\"o}ller}}, \bibinfo {author} {\bibfnamefont {E.}~\bibnamefont {Icking}},
  \bibinfo {author} {\bibfnamefont {K.}~\bibnamefont {Watanabe}}, \bibinfo
  {author} {\bibfnamefont {T.}~\bibnamefont {Taniguchi}}, \bibinfo {author}
  {\bibfnamefont {C.}~\bibnamefont {Volk}},\ and\ \bibinfo {author}
  {\bibfnamefont {C.}~\bibnamefont {Stampfer}},\ }\bibfield  {title} {\bibinfo
  {title} {Single-{{Electron Double Quantum Dots}} in {{Bilayer Graphene}}},\
  }\href {https://doi.org/10.1021/acs.nanolett.9b05295} {\bibfield  {journal}
  {\bibinfo  {journal} {Nano Letters}\ }\textbf {\bibinfo {volume} {20}},\
  \bibinfo {pages} {2005} (\bibinfo {year} {2020}{\natexlab{b}})}\BibitemShut
  {NoStop}%
\bibitem [{\citenamefont {Braem}\ \emph {et~al.}(2018)\citenamefont {Braem},
  \citenamefont {Pellegrino}, \citenamefont {Principi}, \citenamefont
  {R{\"o}{\"o}sli}, \citenamefont {Gold}, \citenamefont {Hennel}, \citenamefont
  {Koski}, \citenamefont {Berl}, \citenamefont {Dietsche}, \citenamefont
  {Wegscheider}, \citenamefont {Polini}, \citenamefont {Ihn},\ and\
  \citenamefont {Ensslin}}]{BraemScanninggatemicroscopy2018}%
  \BibitemOpen
  \bibfield  {author} {\bibinfo {author} {\bibfnamefont {B.~A.}\ \bibnamefont
  {Braem}}, \bibinfo {author} {\bibfnamefont {F.~M.~D.}\ \bibnamefont
  {Pellegrino}}, \bibinfo {author} {\bibfnamefont {A.}~\bibnamefont
  {Principi}}, \bibinfo {author} {\bibfnamefont {M.}~\bibnamefont
  {R{\"o}{\"o}sli}}, \bibinfo {author} {\bibfnamefont {C.}~\bibnamefont
  {Gold}}, \bibinfo {author} {\bibfnamefont {S.}~\bibnamefont {Hennel}},
  \bibinfo {author} {\bibfnamefont {J.~V.}\ \bibnamefont {Koski}}, \bibinfo
  {author} {\bibfnamefont {M.}~\bibnamefont {Berl}}, \bibinfo {author}
  {\bibfnamefont {W.}~\bibnamefont {Dietsche}}, \bibinfo {author}
  {\bibfnamefont {W.}~\bibnamefont {Wegscheider}}, \bibinfo {author}
  {\bibfnamefont {M.}~\bibnamefont {Polini}}, \bibinfo {author} {\bibfnamefont
  {T.}~\bibnamefont {Ihn}},\ and\ \bibinfo {author} {\bibfnamefont
  {K.}~\bibnamefont {Ensslin}},\ }\bibfield  {title} {\bibinfo {title}
  {Scanning gate microscopy in a viscous electron fluid},\ }\href
  {https://doi.org/10.1103/PhysRevB.98.241304} {\bibfield  {journal} {\bibinfo
  {journal} {Physical Review B}\ }\textbf {\bibinfo {volume} {98}},\ \bibinfo
  {pages} {241304} (\bibinfo {year} {2018})}\BibitemShut {NoStop}%
\bibitem [{\citenamefont {Aidala}\ \emph {et~al.}(2007)\citenamefont {Aidala},
  \citenamefont {Parrott}, \citenamefont {Kramer}, \citenamefont {Heller},
  \citenamefont {Westervelt}, \citenamefont {Hanson},\ and\ \citenamefont
  {Gossard}}]{AidalaImagingmagneticfocusing2007}%
  \BibitemOpen
  \bibfield  {author} {\bibinfo {author} {\bibfnamefont {K.~E.}\ \bibnamefont
  {Aidala}}, \bibinfo {author} {\bibfnamefont {R.~E.}\ \bibnamefont {Parrott}},
  \bibinfo {author} {\bibfnamefont {T.}~\bibnamefont {Kramer}}, \bibinfo
  {author} {\bibfnamefont {E.~J.}\ \bibnamefont {Heller}}, \bibinfo {author}
  {\bibfnamefont {R.~M.}\ \bibnamefont {Westervelt}}, \bibinfo {author}
  {\bibfnamefont {M.~P.}\ \bibnamefont {Hanson}},\ and\ \bibinfo {author}
  {\bibfnamefont {A.~C.}\ \bibnamefont {Gossard}},\ }\bibfield  {title}
  {\bibinfo {title} {Imaging magnetic focusing of coherent electron waves},\
  }\href {https://doi.org/10.1038/nphys628} {\bibfield  {journal} {\bibinfo
  {journal} {Nature Physics}\ }\textbf {\bibinfo {volume} {3}},\ \bibinfo
  {pages} {464} (\bibinfo {year} {2007})}\BibitemShut {NoStop}%
\bibitem [{\citenamefont {Bhandari}\ \emph {et~al.}(2016)\citenamefont
  {Bhandari}, \citenamefont {Lee}, \citenamefont {Klales}, \citenamefont
  {Watanabe}, \citenamefont {Taniguchi}, \citenamefont {Heller}, \citenamefont
  {Kim},\ and\ \citenamefont
  {Westervelt}}]{BhandariImagingCyclotronOrbits2016}%
  \BibitemOpen
  \bibfield  {author} {\bibinfo {author} {\bibfnamefont {S.}~\bibnamefont
  {Bhandari}}, \bibinfo {author} {\bibfnamefont {G.-H.}\ \bibnamefont {Lee}},
  \bibinfo {author} {\bibfnamefont {A.}~\bibnamefont {Klales}}, \bibinfo
  {author} {\bibfnamefont {K.}~\bibnamefont {Watanabe}}, \bibinfo {author}
  {\bibfnamefont {T.}~\bibnamefont {Taniguchi}}, \bibinfo {author}
  {\bibfnamefont {E.}~\bibnamefont {Heller}}, \bibinfo {author} {\bibfnamefont
  {P.}~\bibnamefont {Kim}},\ and\ \bibinfo {author} {\bibfnamefont {R.~M.}\
  \bibnamefont {Westervelt}},\ }\bibfield  {title} {\bibinfo {title} {Imaging
  {{Cyclotron Orbits}} of {{Electrons}} in {{Graphene}}},\ }\href
  {https://doi.org/10.1021/acs.nanolett.5b04609} {\bibfield  {journal}
  {\bibinfo  {journal} {Nano Letters}\ }\textbf {\bibinfo {volume} {16}},\
  \bibinfo {pages} {1690} (\bibinfo {year} {2016})}\BibitemShut {NoStop}%
\bibitem [{\citenamefont {Pascher}\ \emph {et~al.}(2014)\citenamefont
  {Pascher}, \citenamefont {R{\"o}ssler}, \citenamefont {Ihn}, \citenamefont
  {Ensslin}, \citenamefont {Reichl},\ and\ \citenamefont
  {Wegscheider}}]{PascherImagingConductanceInteger2014}%
  \BibitemOpen
  \bibfield  {author} {\bibinfo {author} {\bibfnamefont {N.}~\bibnamefont
  {Pascher}}, \bibinfo {author} {\bibfnamefont {C.}~\bibnamefont
  {R{\"o}ssler}}, \bibinfo {author} {\bibfnamefont {T.}~\bibnamefont {Ihn}},
  \bibinfo {author} {\bibfnamefont {K.}~\bibnamefont {Ensslin}}, \bibinfo
  {author} {\bibfnamefont {C.}~\bibnamefont {Reichl}},\ and\ \bibinfo {author}
  {\bibfnamefont {W.}~\bibnamefont {Wegscheider}},\ }\bibfield  {title}
  {\bibinfo {title} {Imaging the {{Conductance}} of {{Integer}} and
  {{Fractional Quantum Hall Edge States}}},\ }\href
  {https://doi.org/10.1103/PhysRevX.4.011014} {\bibfield  {journal} {\bibinfo
  {journal} {Physical Review X}\ }\textbf {\bibinfo {volume} {4}},\ \bibinfo
  {pages} {011014} (\bibinfo {year} {2014})}\BibitemShut {NoStop}%
\bibitem [{\citenamefont {Brun}\ \emph {et~al.}(2014)\citenamefont {Brun},
  \citenamefont {Martins}, \citenamefont {Faniel}, \citenamefont {Hackens},
  \citenamefont {Bachelier}, \citenamefont {Cavanna}, \citenamefont {Ulysse},
  \citenamefont {Ouerghi}, \citenamefont {Gennser}, \citenamefont {Mailly},
  \citenamefont {Huant}, \citenamefont {Bayot}, \citenamefont {Sanquer},\ and\
  \citenamefont {Sellier}}]{BrunWignerKondophysics2014}%
  \BibitemOpen
  \bibfield  {author} {\bibinfo {author} {\bibfnamefont {B.}~\bibnamefont
  {Brun}}, \bibinfo {author} {\bibfnamefont {F.}~\bibnamefont {Martins}},
  \bibinfo {author} {\bibfnamefont {S.}~\bibnamefont {Faniel}}, \bibinfo
  {author} {\bibfnamefont {B.}~\bibnamefont {Hackens}}, \bibinfo {author}
  {\bibfnamefont {G.}~\bibnamefont {Bachelier}}, \bibinfo {author}
  {\bibfnamefont {A.}~\bibnamefont {Cavanna}}, \bibinfo {author} {\bibfnamefont
  {C.}~\bibnamefont {Ulysse}}, \bibinfo {author} {\bibfnamefont
  {A.}~\bibnamefont {Ouerghi}}, \bibinfo {author} {\bibfnamefont
  {U.}~\bibnamefont {Gennser}}, \bibinfo {author} {\bibfnamefont
  {D.}~\bibnamefont {Mailly}}, \bibinfo {author} {\bibfnamefont
  {S.}~\bibnamefont {Huant}}, \bibinfo {author} {\bibfnamefont
  {V.}~\bibnamefont {Bayot}}, \bibinfo {author} {\bibfnamefont
  {M.}~\bibnamefont {Sanquer}},\ and\ \bibinfo {author} {\bibfnamefont
  {H.}~\bibnamefont {Sellier}},\ }\bibfield  {title} {\bibinfo {title} {Wigner
  and {{Kondo}} physics in quantum point contacts revealed by scanning gate
  microscopy},\ }\href {https://doi.org/10.1038/ncomms5290} {\bibfield
  {journal} {\bibinfo  {journal} {Nature Communications}\ }\textbf {\bibinfo
  {volume} {5}},\ \bibinfo {pages} {4290} (\bibinfo {year} {2014})}\BibitemShut
  {NoStop}%
\bibitem [{\citenamefont {Brun}\ \emph {et~al.}(2019)\citenamefont {Brun},
  \citenamefont {Moreau}, \citenamefont {Somanchi}, \citenamefont {Nguyen},
  \citenamefont {Watanabe}, \citenamefont {Taniguchi}, \citenamefont
  {Charlier}, \citenamefont {Stampfer},\ and\ \citenamefont
  {Hackens}}]{BrunImagingDiracfermions2019}%
  \BibitemOpen
  \bibfield  {author} {\bibinfo {author} {\bibfnamefont {B.}~\bibnamefont
  {Brun}}, \bibinfo {author} {\bibfnamefont {N.}~\bibnamefont {Moreau}},
  \bibinfo {author} {\bibfnamefont {S.}~\bibnamefont {Somanchi}}, \bibinfo
  {author} {\bibfnamefont {V.-H.}\ \bibnamefont {Nguyen}}, \bibinfo {author}
  {\bibfnamefont {K.}~\bibnamefont {Watanabe}}, \bibinfo {author}
  {\bibfnamefont {T.}~\bibnamefont {Taniguchi}}, \bibinfo {author}
  {\bibfnamefont {J.-C.}\ \bibnamefont {Charlier}}, \bibinfo {author}
  {\bibfnamefont {C.}~\bibnamefont {Stampfer}},\ and\ \bibinfo {author}
  {\bibfnamefont {B.}~\bibnamefont {Hackens}},\ }\bibfield  {title} {\bibinfo
  {title} {Imaging {{Dirac}} fermions flow through a circular {{Veselago}}
  lens},\ }\href {https://doi.org/10.1103/PhysRevB.100.041401} {\bibfield
  {journal} {\bibinfo  {journal} {Physical Review B}\ }\textbf {\bibinfo
  {volume} {100}},\ \bibinfo {pages} {041401} (\bibinfo {year}
  {2019})}\BibitemShut {NoStop}%
\bibitem [{\citenamefont {Topinka}\ \emph {et~al.}(2000)\citenamefont
  {Topinka}, \citenamefont {LeRoy}, \citenamefont {Shaw}, \citenamefont
  {Heller}, \citenamefont {Westervelt}, \citenamefont {Maranowski},\ and\
  \citenamefont {Gossard}}]{TopinkaImagingCoherentElectron2000}%
  \BibitemOpen
  \bibfield  {author} {\bibinfo {author} {\bibfnamefont {M.~A.}\ \bibnamefont
  {Topinka}}, \bibinfo {author} {\bibfnamefont {B.~J.}\ \bibnamefont {LeRoy}},
  \bibinfo {author} {\bibfnamefont {S.~E.~J.}\ \bibnamefont {Shaw}}, \bibinfo
  {author} {\bibfnamefont {E.~J.}\ \bibnamefont {Heller}}, \bibinfo {author}
  {\bibfnamefont {R.~M.}\ \bibnamefont {Westervelt}}, \bibinfo {author}
  {\bibfnamefont {K.~D.}\ \bibnamefont {Maranowski}},\ and\ \bibinfo {author}
  {\bibfnamefont {A.~C.}\ \bibnamefont {Gossard}},\ }\bibfield  {title}
  {\bibinfo {title} {Imaging {{Coherent Electron Flow}} from a {{Quantum Point
  Contact}}},\ }\href {https://doi.org/10.1126/science.289.5488.2323}
  {\bibfield  {journal} {\bibinfo  {journal} {Science}\ }\textbf {\bibinfo
  {volume} {289}},\ \bibinfo {pages} {2323} (\bibinfo {year}
  {2000})}\BibitemShut {NoStop}%
\bibitem [{\citenamefont {Topinka}\ \emph {et~al.}(2001)\citenamefont
  {Topinka}, \citenamefont {LeRoy}, \citenamefont {Westervelt}, \citenamefont
  {Shaw}, \citenamefont {Fleischmann}, \citenamefont {Heller}, \citenamefont
  {Maranowski},\ and\ \citenamefont
  {Gossard}}]{TopinkaCoherentbranchedflow2001}%
  \BibitemOpen
  \bibfield  {author} {\bibinfo {author} {\bibfnamefont {M.~A.}\ \bibnamefont
  {Topinka}}, \bibinfo {author} {\bibfnamefont {B.~J.}\ \bibnamefont {LeRoy}},
  \bibinfo {author} {\bibfnamefont {R.~M.}\ \bibnamefont {Westervelt}},
  \bibinfo {author} {\bibfnamefont {S.~E.~J.}\ \bibnamefont {Shaw}}, \bibinfo
  {author} {\bibfnamefont {R.}~\bibnamefont {Fleischmann}}, \bibinfo {author}
  {\bibfnamefont {E.~J.}\ \bibnamefont {Heller}}, \bibinfo {author}
  {\bibfnamefont {K.~D.}\ \bibnamefont {Maranowski}},\ and\ \bibinfo {author}
  {\bibfnamefont {A.~C.}\ \bibnamefont {Gossard}},\ }\bibfield  {title}
  {\bibinfo {title} {Coherent branched flow in a two-dimensional electron
  gas},\ }\href {https://doi.org/10.1038/35065553} {\bibfield  {journal}
  {\bibinfo  {journal} {Nature}\ }\textbf {\bibinfo {volume} {410}},\ \bibinfo
  {pages} {183} (\bibinfo {year} {2001})}\BibitemShut {NoStop}%
\bibitem [{\citenamefont {Jura}\ \emph {et~al.}(2007)\citenamefont {Jura},
  \citenamefont {Topinka}, \citenamefont {Urban}, \citenamefont {Yazdani},
  \citenamefont {Shtrikman}, \citenamefont {Pfeiffer}, \citenamefont {West},\
  and\ \citenamefont
  {{Goldhaber-Gordon}}}]{JuraUnexpectedfeaturesbranched2007}%
  \BibitemOpen
  \bibfield  {author} {\bibinfo {author} {\bibfnamefont {M.~P.}\ \bibnamefont
  {Jura}}, \bibinfo {author} {\bibfnamefont {M.~A.}\ \bibnamefont {Topinka}},
  \bibinfo {author} {\bibfnamefont {L.}~\bibnamefont {Urban}}, \bibinfo
  {author} {\bibfnamefont {A.}~\bibnamefont {Yazdani}}, \bibinfo {author}
  {\bibfnamefont {H.}~\bibnamefont {Shtrikman}}, \bibinfo {author}
  {\bibfnamefont {L.~N.}\ \bibnamefont {Pfeiffer}}, \bibinfo {author}
  {\bibfnamefont {K.~W.}\ \bibnamefont {West}},\ and\ \bibinfo {author}
  {\bibfnamefont {D.}~\bibnamefont {{Goldhaber-Gordon}}},\ }\bibfield  {title}
  {\bibinfo {title} {Unexpected features of branched flow through high-mobility
  two-dimensional electron gases},\ }\href {https://doi.org/10.1038/nphys756}
  {\bibfield  {journal} {\bibinfo  {journal} {Nature Physics}\ }\textbf
  {\bibinfo {volume} {3}},\ \bibinfo {pages} {841} (\bibinfo {year}
  {2007})}\BibitemShut {NoStop}%
\bibitem [{\citenamefont {Zhang}\ \emph {et~al.}(2009)\citenamefont {Zhang},
  \citenamefont {Tang}, \citenamefont {Girit}, \citenamefont {Hao},
  \citenamefont {Martin}, \citenamefont {Zettl}, \citenamefont {Crommie},
  \citenamefont {Shen},\ and\ \citenamefont
  {Wang}}]{ZhangDirectobservationwidely2009}%
  \BibitemOpen
  \bibfield  {author} {\bibinfo {author} {\bibfnamefont {Y.}~\bibnamefont
  {Zhang}}, \bibinfo {author} {\bibfnamefont {T.-T.}\ \bibnamefont {Tang}},
  \bibinfo {author} {\bibfnamefont {C.}~\bibnamefont {Girit}}, \bibinfo
  {author} {\bibfnamefont {Z.}~\bibnamefont {Hao}}, \bibinfo {author}
  {\bibfnamefont {M.~C.}\ \bibnamefont {Martin}}, \bibinfo {author}
  {\bibfnamefont {A.}~\bibnamefont {Zettl}}, \bibinfo {author} {\bibfnamefont
  {M.~F.}\ \bibnamefont {Crommie}}, \bibinfo {author} {\bibfnamefont {Y.~R.}\
  \bibnamefont {Shen}},\ and\ \bibinfo {author} {\bibfnamefont
  {F.}~\bibnamefont {Wang}},\ }\bibfield  {title} {\bibinfo {title} {Direct
  observation of a widely tunable bandgap in bilayer graphene},\ }\href
  {https://doi.org/10.1038/nature08105} {\bibfield  {journal} {\bibinfo
  {journal} {Nature}\ }\textbf {\bibinfo {volume} {459}},\ \bibinfo {pages}
  {820} (\bibinfo {year} {2009})}\BibitemShut {NoStop}%
\bibitem [{Note1()}]{Note1}%
  \BibitemOpen
  \bibinfo {note} {The upper minimum most likely arises due to the charge
  neutrality point of the bulk regions close to the channel. The additional,
  second minimum occurs due to another device in parallel to the channel, which
  is not depicted in the schematic in Fig. \ref {fig: Fig1}a and was grounded
  during the measurement.}\BibitemShut {Stop}%
\bibitem [{\citenamefont {Gold}\ \emph {et~al.}(2020)\citenamefont {Gold},
  \citenamefont {Kurzmann}, \citenamefont {Watanabe}, \citenamefont
  {Taniguchi}, \citenamefont {Ensslin},\ and\ \citenamefont
  {Ihn}}]{GoldScanningGateMicroscopy2020}%
  \BibitemOpen
  \bibfield  {author} {\bibinfo {author} {\bibfnamefont {C.}~\bibnamefont
  {Gold}}, \bibinfo {author} {\bibfnamefont {A.}~\bibnamefont {Kurzmann}},
  \bibinfo {author} {\bibfnamefont {K.}~\bibnamefont {Watanabe}}, \bibinfo
  {author} {\bibfnamefont {T.}~\bibnamefont {Taniguchi}}, \bibinfo {author}
  {\bibfnamefont {K.}~\bibnamefont {Ensslin}},\ and\ \bibinfo {author}
  {\bibfnamefont {T.}~\bibnamefont {Ihn}},\ }\bibfield  {title} {\bibinfo
  {title} {Scanning {{Gate Microscopy}} of {{Localized States}} in a
  gate-defined {{Bilayer Graphene Channel}}},\ }\href@noop {} {\bibfield
  {journal} {\bibinfo  {journal} {arXiv:2006.10144 [cond-mat]}\ } (\bibinfo
  {year} {2020})},\ \Eprint {https://arxiv.org/abs/2006.10144}
  {arXiv:2006.10144 [cond-mat]} \BibitemShut {NoStop}%
\bibitem [{\citenamefont {P{\'e}terfalvi}\ \emph {et~al.}(2012)\citenamefont
  {P{\'e}terfalvi}, \citenamefont {Oroszl{\'a}ny}, \citenamefont {Lambert},\
  and\ \citenamefont {Cserti}}]{PeterfalviIntrabandelectronfocusing2012}%
  \BibitemOpen
  \bibfield  {author} {\bibinfo {author} {\bibfnamefont {C.~G.}\ \bibnamefont
  {P{\'e}terfalvi}}, \bibinfo {author} {\bibfnamefont {L.}~\bibnamefont
  {Oroszl{\'a}ny}}, \bibinfo {author} {\bibfnamefont {C.~J.}\ \bibnamefont
  {Lambert}},\ and\ \bibinfo {author} {\bibfnamefont {J.}~\bibnamefont
  {Cserti}},\ }\bibfield  {title} {\bibinfo {title} {Intraband electron
  focusing in bilayer graphene},\ }\href
  {https://doi.org/10.1088/1367-2630/14/6/063028} {\bibfield  {journal}
  {\bibinfo  {journal} {New Journal of Physics}\ }\textbf {\bibinfo {volume}
  {14}},\ \bibinfo {pages} {063028} (\bibinfo {year} {2012})}\BibitemShut
  {NoStop}%
\bibitem [{\citenamefont {McCann}\ \emph {et~al.}(2007)\citenamefont {McCann},
  \citenamefont {Abergel},\ and\ \citenamefont
  {Fal'ko}}]{McCannlowenergyelectronic2007}%
  \BibitemOpen
  \bibfield  {author} {\bibinfo {author} {\bibfnamefont {E.}~\bibnamefont
  {McCann}}, \bibinfo {author} {\bibfnamefont {D.~S.}\ \bibnamefont
  {Abergel}},\ and\ \bibinfo {author} {\bibfnamefont {V.~I.}\ \bibnamefont
  {Fal'ko}},\ }\bibfield  {title} {\bibinfo {title} {The low energy electronic
  band structure of bilayer graphene},\ }\href
  {https://doi.org/10.1140/epjst/e2007-00229-1} {\bibfield  {journal} {\bibinfo
   {journal} {The European Physical Journal Special Topics}\ }\textbf {\bibinfo
  {volume} {148}},\ \bibinfo {pages} {91} (\bibinfo {year} {2007})}\BibitemShut
  {NoStop}%
\bibitem [{\citenamefont {Slizovskiy}\ \emph {et~al.}(2019)\citenamefont
  {Slizovskiy}, \citenamefont {{Garcia-Ruiz}}, \citenamefont {Drummond},\ and\
  \citenamefont {Falko}}]{SlizovskiyDielectricsusceptibilitygraphene2019}%
  \BibitemOpen
  \bibfield  {author} {\bibinfo {author} {\bibfnamefont {S.}~\bibnamefont
  {Slizovskiy}}, \bibinfo {author} {\bibfnamefont {A.}~\bibnamefont
  {{Garcia-Ruiz}}}, \bibinfo {author} {\bibfnamefont {N.}~\bibnamefont
  {Drummond}},\ and\ \bibinfo {author} {\bibfnamefont {V.~I.}\ \bibnamefont
  {Falko}},\ }\bibfield  {title} {\bibinfo {title} {Dielectric susceptibility
  of graphene describing its out-of-plane polarizability},\ }\href@noop {}
  {\bibfield  {journal} {\bibinfo  {journal} {arXiv:1912.10067 [cond-mat]}\ }
  (\bibinfo {year} {2019})},\ \Eprint {https://arxiv.org/abs/1912.10067}
  {arXiv:1912.10067 [cond-mat]} \BibitemShut {NoStop}%
\bibitem [{\citenamefont {Overweg}\ \emph
  {et~al.}(2018{\natexlab{b}})\citenamefont {Overweg}, \citenamefont {Knothe},
  \citenamefont {Fabian}, \citenamefont {Linhart}, \citenamefont {Rickhaus},
  \citenamefont {Wernli}, \citenamefont {Watanabe}, \citenamefont {Taniguchi},
  \citenamefont {S{\'a}nchez}, \citenamefont {Burgd{\"o}rfer}, \citenamefont
  {Libisch}, \citenamefont {Fal'ko}, \citenamefont {Ensslin},\ and\
  \citenamefont {Ihn}}]{OverwegTopologicallyNontrivialValley2018}%
  \BibitemOpen
  \bibfield  {author} {\bibinfo {author} {\bibfnamefont {H.}~\bibnamefont
  {Overweg}}, \bibinfo {author} {\bibfnamefont {A.}~\bibnamefont {Knothe}},
  \bibinfo {author} {\bibfnamefont {T.}~\bibnamefont {Fabian}}, \bibinfo
  {author} {\bibfnamefont {L.}~\bibnamefont {Linhart}}, \bibinfo {author}
  {\bibfnamefont {P.}~\bibnamefont {Rickhaus}}, \bibinfo {author}
  {\bibfnamefont {L.}~\bibnamefont {Wernli}}, \bibinfo {author} {\bibfnamefont
  {K.}~\bibnamefont {Watanabe}}, \bibinfo {author} {\bibfnamefont
  {T.}~\bibnamefont {Taniguchi}}, \bibinfo {author} {\bibfnamefont
  {D.}~\bibnamefont {S{\'a}nchez}}, \bibinfo {author} {\bibfnamefont
  {J.}~\bibnamefont {Burgd{\"o}rfer}}, \bibinfo {author} {\bibfnamefont
  {F.}~\bibnamefont {Libisch}}, \bibinfo {author} {\bibfnamefont {V.~I.}\
  \bibnamefont {Fal'ko}}, \bibinfo {author} {\bibfnamefont {K.}~\bibnamefont
  {Ensslin}},\ and\ \bibinfo {author} {\bibfnamefont {T.}~\bibnamefont {Ihn}},\
  }\bibfield  {title} {\bibinfo {title} {Topologically {{Nontrivial Valley
  States}} in {{Bilayer Graphene Quantum Point Contacts}}},\ }\href
  {https://doi.org/10.1103/PhysRevLett.121.257702} {\bibfield  {journal}
  {\bibinfo  {journal} {Physical Review Letters}\ }\textbf {\bibinfo {volume}
  {121}},\ \bibinfo {pages} {257702} (\bibinfo {year}
  {2018}{\natexlab{b}})}\BibitemShut {NoStop}%
\bibitem [{\citenamefont {Knothe}\ and\ \citenamefont
  {Fal'ko}(2018)}]{KnotheInfluenceminivalleysBerry2018}%
  \BibitemOpen
  \bibfield  {author} {\bibinfo {author} {\bibfnamefont {A.}~\bibnamefont
  {Knothe}}\ and\ \bibinfo {author} {\bibfnamefont {V.}~\bibnamefont
  {Fal'ko}},\ }\bibfield  {title} {\bibinfo {title} {Influence of minivalleys
  and {{Berry}} curvature on electrostatically induced quantum wires in gapped
  bilayer graphene},\ }\href {https://doi.org/10.1103/PhysRevB.98.155435}
  {\bibfield  {journal} {\bibinfo  {journal} {Physical Review B}\ }\textbf
  {\bibinfo {volume} {98}},\ \bibinfo {pages} {155435} (\bibinfo {year}
  {2018})}\BibitemShut {NoStop}%
\bibitem [{\citenamefont {Lane}\ \emph {et~al.}(2019)\citenamefont {Lane},
  \citenamefont {Knothe},\ and\ \citenamefont
  {Fal'ko}}]{LaneSemimetallicfeaturesquantum2019b}%
  \BibitemOpen
  \bibfield  {author} {\bibinfo {author} {\bibfnamefont {T.~L.~M.}\
  \bibnamefont {Lane}}, \bibinfo {author} {\bibfnamefont {A.}~\bibnamefont
  {Knothe}},\ and\ \bibinfo {author} {\bibfnamefont {V.~I.}\ \bibnamefont
  {Fal'ko}},\ }\bibfield  {title} {\bibinfo {title} {Semimetallic features in
  quantum transport through a gate-defined point contact in bilayer graphene},\
  }\href {https://doi.org/10.1103/PhysRevB.100.115427} {\bibfield  {journal}
  {\bibinfo  {journal} {Physical Review B}\ }\textbf {\bibinfo {volume}
  {100}},\ \bibinfo {pages} {115427} (\bibinfo {year} {2019})}\BibitemShut
  {NoStop}%
\end{thebibliography}%

\end{document}